\documentclass[aps,pra,twocolumn,amsmath,amssymb,showpacs,preprintnumbers]{revtex4}

\usepackage{dcolumn}
\usepackage{graphicx}
\usepackage{bm}
\input epsf

\begin{document}

\bibliographystyle{unsrt}

\preprint{}

\title{Electron EDM searches based on alkali or alkaline earth bearing molecules}
\author{Edmund R. Meyer}
\email{meyere@murphy.colorado.edu}
\affiliation{JILA, NIST and University of Colorado, Department of Physics,
  Boulder, Colorado 80309-0440, USA}
\author{John L. Bohn}
\affiliation{JILA, NIST and University of Colorado, Department of Physics,
  Boulder, Colorado 80309-0440, USA}
\date{\today}

\begin{abstract}
We introduce four new molecules --- YbRb, YbCs, YbSr$^+$, and YbBa$^+$ --- 
that may prove fruitful in experimental searches for the electric dipole 
moment (EDM) of the electron. These molecules can, in principle, be prepared 
at extremely low temperatures by photoassociating ultracold atoms, and 
therefore may present an advantage over molecular beam experiments. Here 
we discuss properties of these molecules and assess the effective electric 
fields they contribute to an electron EDM measurement.
\end{abstract}


\maketitle
\section{Introduction}

It is by now widely acknowledged that experiments performed on atoms
and molecules afford the best chance of measuring, or else limiting
the possible size of, the electric dipole moment (EDM) of the electron
\cite{suzuki,eedm1,eedm2}.  A lone valence electron inside an atom with 
large nuclear charge $Z$ will feel an effective electric field enhanced 
beyond the field that is applied.  For example, in atomic cesium the apparent
field is 132$\times{\mathcal E}_{\rm app}$, where ${\mathcal E}_{\rm app}$ 
is the applied electric field\cite{krippy}. This enhancement arises from the 
field's ability to mix the opposite-parity $6s$ and $6p$ states of Cs, and a 
fairly large applied electric field is still required. More recent 
estimates using {\it ab initio} methods give enhancement factors for 
Rb and Cs of 25.74 and 120.53 respectively\cite{paraatoms}. Experiments 
on the Rb\cite{rbedm} and Cs\cite{csedm} atoms are currently being pursued 
to take advantage of the enhancement factor and long hold times achievable 
in the alkali atom family.

Thus, to achieve effective electric fields of about 1--2~GV/cm one would 
have to apply fields of $\sim$10--40~MV/cm to these atoms. This is not 
desirable in practice. However, one can hold ultracold Rb and Cs atoms in 
traps for a time long compared to the typical molecular beam experiment. 
Therefore, one applies a smaller electric field while taking advantage of 
the long coherence times. It is the trade-off between hold time and effective 
electric field size that makes atomic electron EDM experiments attractive 
and worth pursuing.

Herein lies the rub: one wants a very large applied electric field,
but with great electric fields come great systematic problems. It
would therefore be beneficial if the $s$-$p$ mixing could be achieved without 
the application of large laboratory electric fields. This is where molecules 
become interesting, since they can be quite adept at mixing $s$- and $p$-atomic 
orbitals on their own, without the need for large, applied electric fields. The 
field requirements to polarize the molecule may then be fairly modest. This 
important insight was initially due to Sandars \cite{ColSandars}.

Typically, experiments aimed at measuring the electron EDM in molecules have
considered the simplest case of molecules with ${}^2\Sigma$
symmetry. Then there is only one unpaired electronic spin and it has
zero orbital angular momentum about the internuclear axis. One
appealing way to achieve fairly robust molecules for this purpose is
to pair an atom in an $s^2$ configuration (such as Yb or Hg)  with
fluorine. Fluorine tends to capture an electron from any atom it
encounters. Thus, when one of the s-electrons is captured by F, a
${}^2\Sigma$ molecule is formed, as in YbF\cite{eedm1}. Moreover, the 
largely ionic bond is responsible for a large effective internal electric 
field that efficiently mixes the atomic $s$- and $p$-orbitals of the heavy atom 
\cite{eedm1,BaF}.

In this article we suggest another class of $^2 \Sigma$ molecules
for EDM searches.  Our emphasis is to consider molecules composed of
atoms that can be laser cooled, thus greatly reducing the molecular
temperature well below that available in molecular beam experiments.
Specifically, we consider the neutral molecules YbRb and YbCs, along
with the molecular ions YbSr$^+$ and YbBa$^+$.  In all cases, laser
cooling of the neutral atoms has been demonstrated \cite{yblc,balc,srlc}. 
Indeed, recently short-lived ultracold samples of metastable YbRb$^\star$ 
have been produced via photoassociation \cite{Gorlitz}. This idea is very 
similar to the proposed van der Waals molecule CsXe\cite{CsXe}.

These molecules, all of $^2 \Sigma$ symmetry in their ground states,
will have both advantages and disadvantages with respect to other
molecules that are considered in EDM searches. The chief
disadvantage is that their electronic wave function suggests a sharing, 
or co-valency, of the bond and this somewhat limits the size of the effective 
electric field that they can sustain. We find that these fields are typically 
far smaller ($10$--$100$ times) than the leading contender --- ThO, with an 
${\cal E}_{\rm eff} = 104$~GV/cm\cite{meyer08}. Their main advantage is that
they are cold enough to be trapped in either optical dipole traps or
ion traps, leading conceivably to large coherence times within which
to perform the experiment.  In addition, these molecules, consisting
of two heavy atoms, will have smaller rotational constants than the
fluorides, hence can be polarized in somewhat smaller fields. This
is a feature that can help reduce systematic effects. Spin-rotational 
effects will also be reduced since they scale inversely with the reduced 
mass of the system; the electron remains fairly well decoupled from the 
molecular axis in the lower-lying rotational levels of the molecules. 
Another novel --- though not necessarily beneficial --- feature is that both 
atoms contain large-$Z$ nuclei and thus can each contribute to the effective 
field. 

In this article we therefore estimate the basic properties of these
molecules, especially the effective electric fields they harbor.
Sec.~II will discuss the YbRb and YbCs molecules and give the details
of the potential energy surfaces and molecular parameters of the
ground and excited states of interest. Sec.~III will present the
same information for the YbSr$^+$ and YbBa$^+$ molecular ions, as
well as the neutrals from which they are derived. Sec.~IV will
detail the effective electric field calculations on the ${}^2\Sigma$
states of interest. Sec.~V will present some concluding comments.

\section{Alkali-Ytterbium molecules}
\label{s:alkali}

Alkali atoms such as Rb and Cs are the workhorses of laser cooling.
\cite{Metcalf}. It is the unpaired $s$-electron that is of interest in an 
electron EDM search since it is what we hope to coax into an $s$-$p$ 
hybridized atomic state. In the introduction we discussed the role of fluorine,
which can extract an electron from a closed $s$-shell atom such as Ba or Yb. 
We can accomplish the same feat by pairing an alkali atom such as Rb with 
a closed $s$-shell atom such as Yb because Yb is more electronegative than 
any alkali save H. This is similar to the pairing of Cs with Xe, as in 
\cite{CsXe}, except Xe is a closed $p$-shell atom. Xe serves as a polarizing agent 
on the Cs atom, causing the Cs $s$- and $p$-orbitals to hybridize. Moreover, the 
difference in electro-negativity is greater between Cs and Yb than between Rb and 
Yb, so that the YbCs molecule might be expected to be the more polar of the two. 
However, it is also important to consider the bonding length. A larger bond 
length would work to counteract the charge transfer mechanism because the two 
atoms would not get close enough together to allow the atomic wave functions to
overlap. In order to understand the true details we turn to {\it ab initio} 
methods in order to gain understanding into the nature of the bonding process.

For the YbRb and YbCs systems we use the {\sc molpro} suite of codes
\cite{MOLPRO}. We start at an inter-atomic distance $R=25\;a_{\rm 0}$ and 
move in to $R=4.8\;a_{\rm 0}$. At $R$=25\;$a_{\rm 0}$ we perform a self 
consistent field Hartree-Fock (SCF-HF) calculation on the closed shell 
Yb-alkali molecular ion system to obtain a starting wave function. We then 
perform a second SCF-HF on the neutral YbRb molecule followed by a 
multi-configuration self consistent field calculation (MCSCF) 
\cite{mcscf1,mcscf2}. The MCSCF contains the ground $X{}^2\Sigma$ as well 
as the excited $a{}^2\Pi$ and $b{}^2\Sigma$ states arising from the alkali 
$n\,p$ configurations. This is the starting configuration for the multi 
reference configuration interaction (MRCI) \cite{mrci1,mrci2} calculation 
performed in each molecular symmetry group. We use an active space of 
\{6,3,3,0\} with no closed orbitals. We used the Stuttgart basis sets and 
effective core potentials (ECP) with core polarization potentials (CPP) for the 
alkali and Yb. We used the ECP68MDF potential \cite{stutYb} for Yb, ECP36SDF 
potential for Rb \cite{stutRb,stutRb2} and ECP54SDF potential for Cs\cite{stutRb}. 
All of the basis sets associated with the ECPs were left uncontracted. The active 
spaces in the calculations were chosen such that the experimental excitation 
energy of the alkali was reproduced to within 50\;cm$^{\rm -1}$.

\begin{figure}\label{figs1ab}
  \begin{center}
    \resizebox{3.25in}{!}{\includegraphics{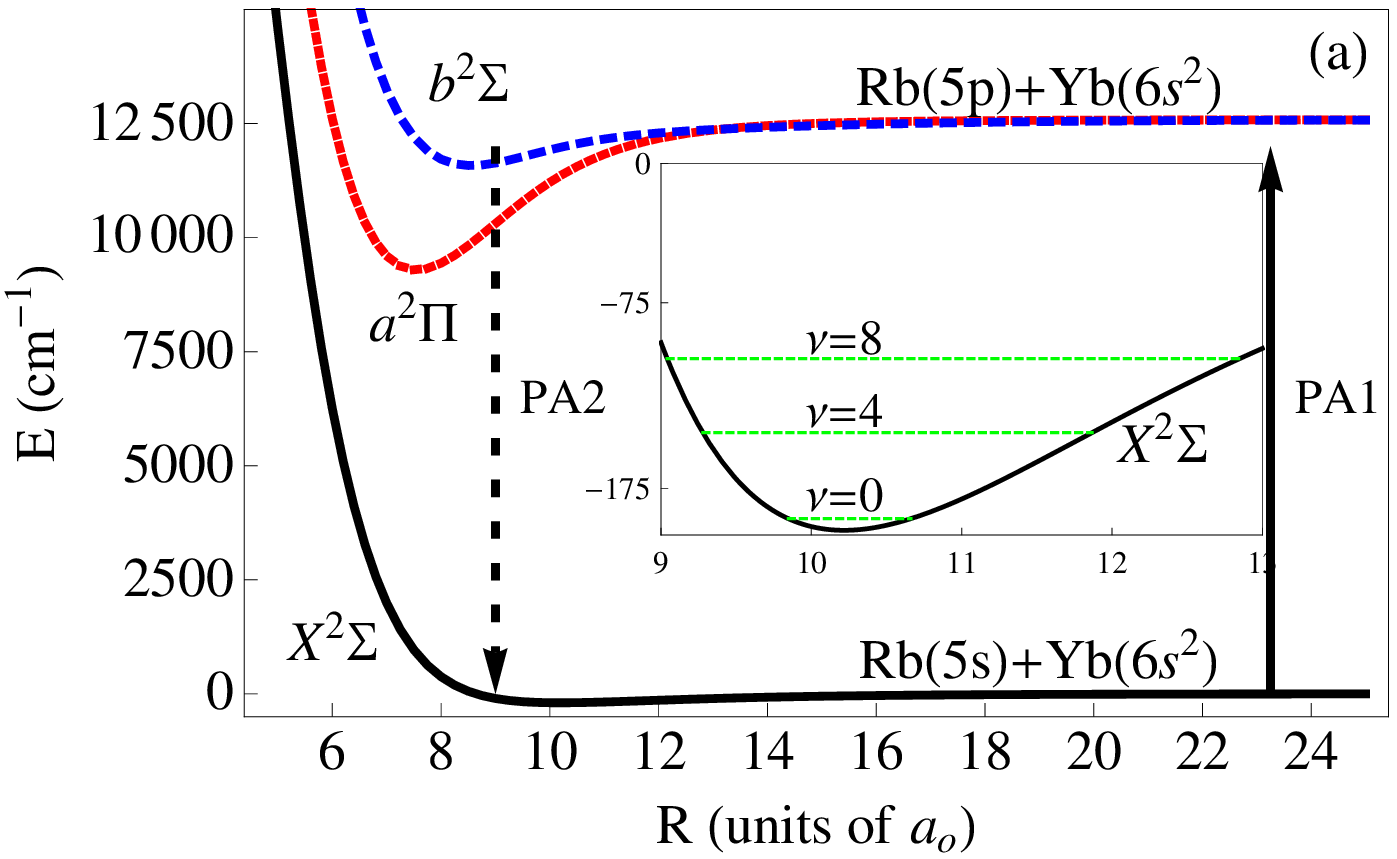}}
    \resizebox{3.25in}{!}{\includegraphics{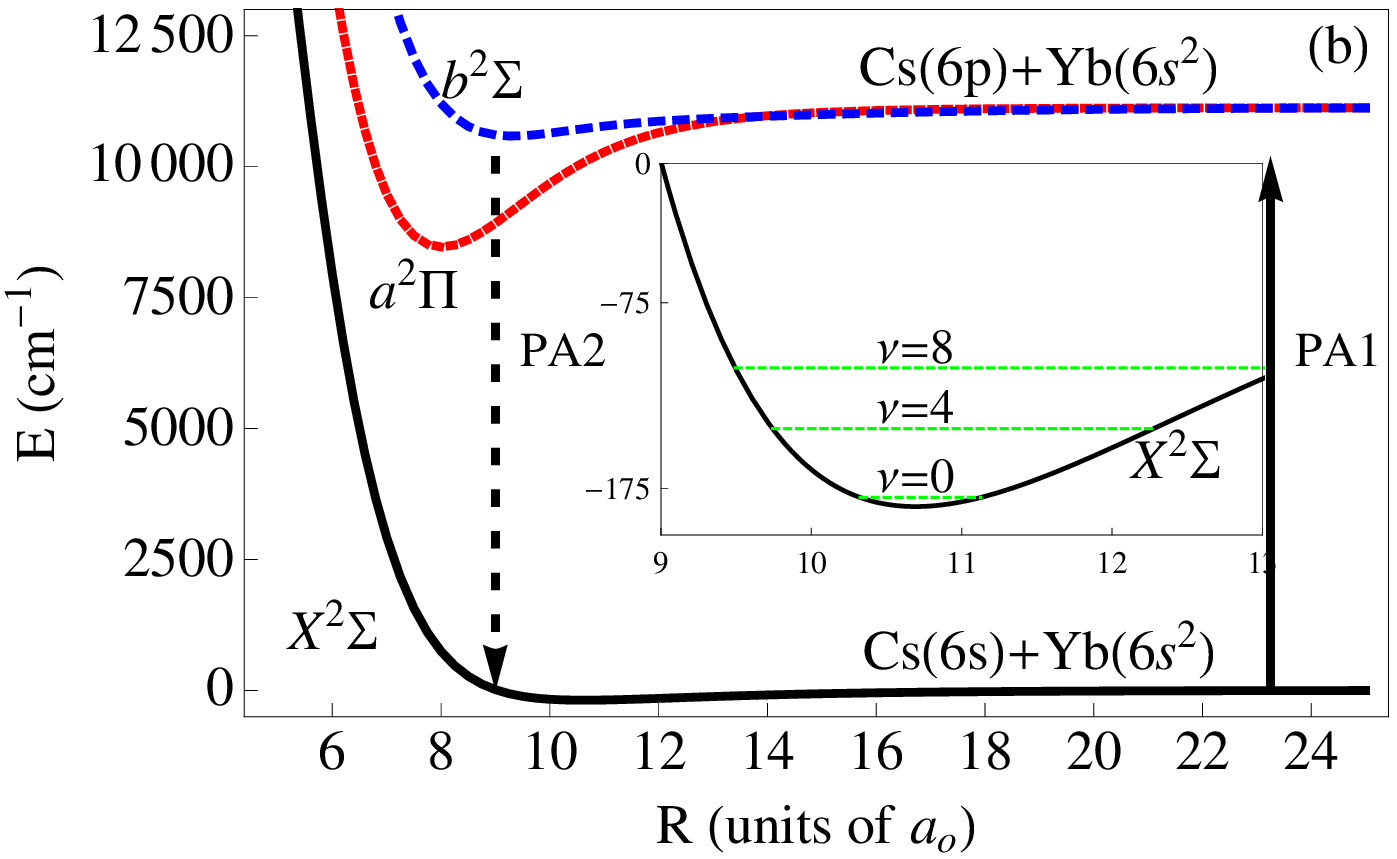}}
  \end{center}
  \caption{(Color online) Potential energy surfaces for the YbRb (a) and YbCs 
    (b) molecules. $a_o$ is the Bohr length and is 0.52918 $\AA$. The ground 
    to excited state energy splitting in the separated limit was optimized to 
    produce the Rb (Cs) $5(6)\;s\rightarrow5(6)\;p$ transition known 
    experimentally. The arrows indicate the two-color photoassociation route 
    to producing ultracold polar molecules in the ground $X{}^2\Sigma$ state. 
    The inset shows the ground vibrational levels $\nu=0,4,8$.}
\end{figure}

The results of the potential calculations are presented in
Figs.~\ref{figs1ab} a and b, for the ground state and the two lowest-lying
dipole-allowed excited states in Hund's coupling case (a).  The
inset is an enlargement of the ground state showing that, while
far shallower than alkali dimer potentials, it is expected to hold
bound states. Table~\ref{tab1} gives the relevant molecular
parameters such as bond length, vibrational constant, dissociation
energy, and rotation constant. The parameters were obtained by
fitting the {\it ab initio} points to a Morse potential. Both YbCs
and YbRb exhibit a large bonding length greater than 10\;$a_{\rm o}$
in the ground $X{}^2\Sigma$ state. Due to the very shallow potential and 
large bond length, these molecules appear to be of a van der Waals-type 
bonding nature, similar to that found in CsXe\cite{CsXe}.

In each figure the solid arrow denotes the photoassociation step
needed to drive the initial atom pair into an excited state
molecule. For YbRb, this step has already been demonstrated
experimentally \cite{Gorlitz}. Also shown schematically as a dashed arrow is
the transition necessary to create molecules in the ground
electronic, vibrational and rotational state.  Depending on which
levels can ultimately be accessed in the first photoassociation
step, this second transition may not be possible in a single step.
The quest to produce ultracold, ground state polar molecules from 
ultracold atoms is currently a major research area, which has recently 
seen the production of molecular ground states\cite{jinye,demhud}.

\begin{table}
  \caption{\label{tab1} Molecular parameters for YbRb and YbCs. Bond distance
    $r_e$, dissociation energy ($D$), vibration constants ($\omega_e$ and
    $\omega_e\chi_e$), rotation constants ($B_e$ and $D_e$), and 
    vibration-rotation mixing constants $\alpha_e$. $r_e$ is in atomic units 
    while all others are in cm$^{\rm -1}$.}
  \begin{ruledtabular}
    \begin{tabular}{l|l|l|l|l|l|l|l}
      Molecule & $r_e$ & $D$ & $\omega_e$ & $\omega_e\chi_e$ & 
      $B_e$\,$10^{-2}$ & $D_e$\,$10^{-9}$ & $\alpha_e$\,$10^{-4}$ \\
      \hline
      YbRb $X{}^2\Sigma$& 10.22 & 194 & 12.6 & 0.21 & 0.99 & 24 & 1.65\\
      YbRb $a{}^2\Pi$ & 7.44 & 3380 & 66.4 & 0.33 & 1.86 & 5.9 & 1.00\\
      YbRb $b{}^2\Sigma$ & 8.52 & 942 & 36.0 & 0.34 & 1.42 & 8.8 & 1.32\\
      \hline
      YbCs $X{}^2\Sigma$& 10.69 & 182 & 9.9 & 0.13 & 0.69 & 14 & 0.99\\
      YbCs $a{}^2\Pi$ & 7.91 & 2710 & 50.9 & 0.24 & 1.27 & 3.2 & 0.63\\
      YbCs $b{}^2\Sigma$ & 9.41 & 490 & 20.9 & 0.22 & 0.90 & 6.6 & 092
    \end{tabular}
  \end{ruledtabular}
\end{table}

The {\it ab initio} calculations show that while the dominant contribution 
to the unpaired electron is from the Rb $5s$ (Cs $6s$) orbital, there are 
appreciable contributions from the Rb $5p$ (Cs $6p$) as well as the Yb 
$6s$ and $6p$ atomic configurations. A secondary contribution comes from 
a similar arrangement except that the Yb $6p$ atomic orbital has a more 
pronounced presence in the second $\sigma$-molecular orbital. There is some, 
albeit little, charge transfer from the alkali to Yb. At the potential minima, 
Rb transfers more charge to Yb than does Cs, counter to what is expected based 
on electro-negativity arguments. However, this can be understood in terms of 
the bond length in YbCs being larger than in YbRb. The excited $a{}^2\Pi$ state 
of both YbRb and YbCs contains a far shorter bond length as well as a deeper well 
than the ground state $X{}^2\Sigma$ state. This is due to the larger 
polarizability in the $5(6)\;p$ state of Rb (Cs).

\begin{figure}\label{fig2}
  \begin{center}
    \resizebox{3.25in}{!}{\includegraphics{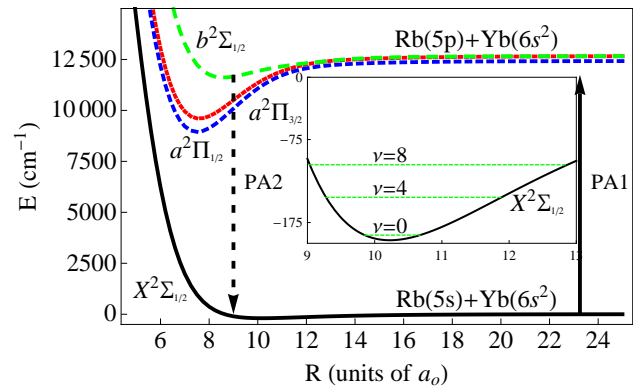}}
  \end{center}
  \caption{(Color online) YbRb potential energy curves with spin-orbit 
    effects included. $a_0$ is the Bohr length, $0.52918~\AA$. The
    effective core potential ECP36SDF was modified to produce the
    $5p{}^2P_{1/2}\rightarrow5p{}^2P_{3/2}$ energy splitting of Rb. 
    The arrows indicate the two-color photoassociation route. The inset 
    shows the ground vibrational levels $\nu=0,4,8$.}
\end{figure}

In order to obtain a higher level of accuracy in describing the
excited $a{}^2\Pi$ state, we included the effects of spin-orbit
coupling in the {\it ab initio} calculations. We then modified our
long range potential to match the experimentally determined $C_6$
coefficient determined by \cite{Gorlitz}. Fig.~\ref{fig2}
gives the potentials with spin-orbit terms applied throughout. The
spin-orbit terms of the ECP36SDF effective core potential of the
Stuttgart group were modified so as to reproduce the
$5p{}^2P_{1/2}\rightarrow5p{}^2P_{3/2}$ energy splitting of the Rb
atom to within a few cm$^{\rm -1}$. The $X{}^2\Sigma_{1/2}$ state is
fairly unaffected by the presence of the spin-orbit terms; this is due to
the large electronic energy separation of the ground and excited
electronic states compared to the atomic spin-orbit constant in the 
alkali atoms. There are now 3 excited states since the ${}^2\Pi$ state 
splits into two surfaces, one each for where the electronic
spin lies parallel or anti-parallel to the electronic orbital
angular momentum. In addition, at large separations of about
11\;$a_{\rm o}$ there is an avoided crossing between the the
$a{}^2\Pi_{1/2}$ and the $b{}^2\Sigma_{1/2}$ states due to the
spin-orbit interaction. This changes the outer wall behavior of 
both states.

We again fit the {\it ab initio} points to a Morse potential in order to
obtain the spectroscopic parameters describing the states of interest --- 
see Table \ref{tab3}. There is very little change in the parameters for the
$X{}^2\Sigma_{1/2}$ state. The excited $a{}^2\Pi_{1/2}$ state has a slightly
smaller bond length but larger well depth while the the $a{}^2\Pi_{3/2}$ state
has a longer bond length and smaller well depth compared to the $a{}^2\Pi$ 
state in the previous calculation.
\begin{table}
  \caption{\label{tab3} Molecular parameters for YbRb including spin-orbital 
    effects. Bond distance $r_e$, dissociation energy ($D$), vibration constants 
    ($\omega_e$ and $\omega_e\chi_e$), rotation constants ($B_e$ and $D_e$), 
    and vibration-rotation mixing constant $\alpha_e$ are all determined from 
    the Morse potential parameter. $r_e$ is in atomic units while all others are 
    in cm$^{\rm -1}$.}
  \begin{ruledtabular}
    \begin{tabular}{l|l|l|l|l|l|l|l}
      Molecule & $r_e$ & $D$ & $\omega_e$ & $\omega_e\chi_e$ & 
      $B_e$\,$10^{-2}$ & $D_e$\,$10^{-9}$ & $\alpha_e$\,$10^{-4}$ \\
      \hline
      YbRb $X{}^2\Sigma_{1/2}$& 10.23 & 193 & 12.4 & 0.20 & 0.99 & 25 & 1.64\\
      YbRb $a{}^2\Pi_{1/2}$ & 7.39 & 3590 & 70.9 & 0.35 & 1.89 & 5.4 & 1.00\\
      YbRb $a{}^2\Pi_{3/2}$ & 7.52 & 3120 & 65.4 & 0.34 & 1.83 & 5.7 & 1.02\\
      YbRb $b{}^2\Sigma_{1/2}$ & 8.58 & 1026 & 33.9 & 0.28 & 1.40 & 9.6 & 1.20
    \end{tabular}
  \end{ruledtabular}
\end{table}

\section{Alkaline Earth-Ytterbium Molecular Ions}

In addition to the neutral molecules considered in the previous section, 
we can also consider their iso-electronic cation partners YbSr$^+$ 
and YbBa$^+$. The molecules might be expected to form deeper wells and 
shorter bond lengths because the atomic cations Sr and Ba will strongly 
attract an electron. Also, these ions offer intriguing possibilities for an EDM
experiment in a stable ion trap, as has been recently proposed\cite{eric}.
Another possibility is to arrange the molecular ions into an optical 
lattice. In addition, replacing the alkali atoms with alkaline earths
afford the possibility of isotopomers with zero nuclear spin on both
atoms, thus greatly simplifying their spectroscopy.

\begin{figure}\label{fig3}
  \begin{center}
    \resizebox{3.25in}{!}{\includegraphics{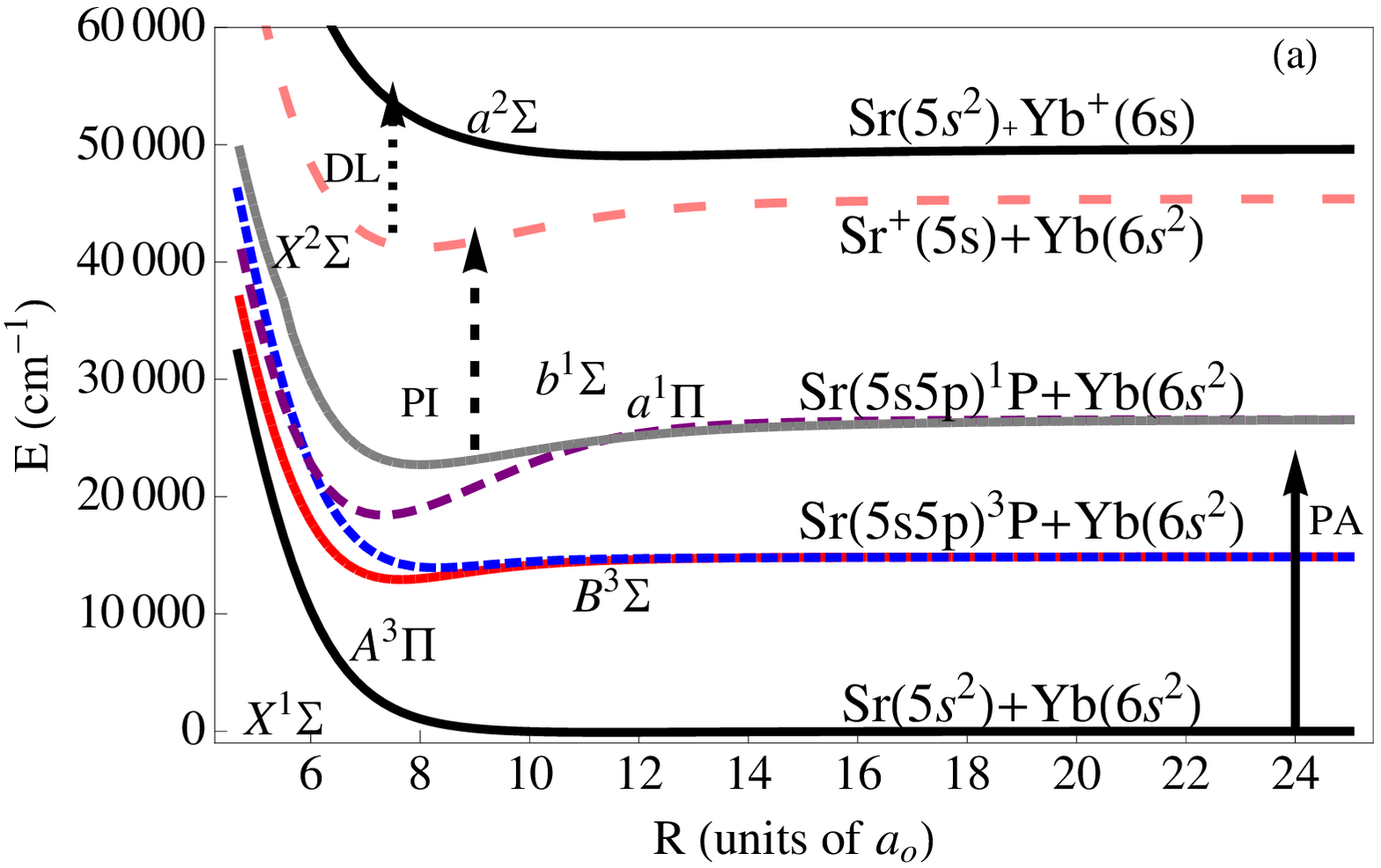}}
    \resizebox{3.25in}{!}{\includegraphics{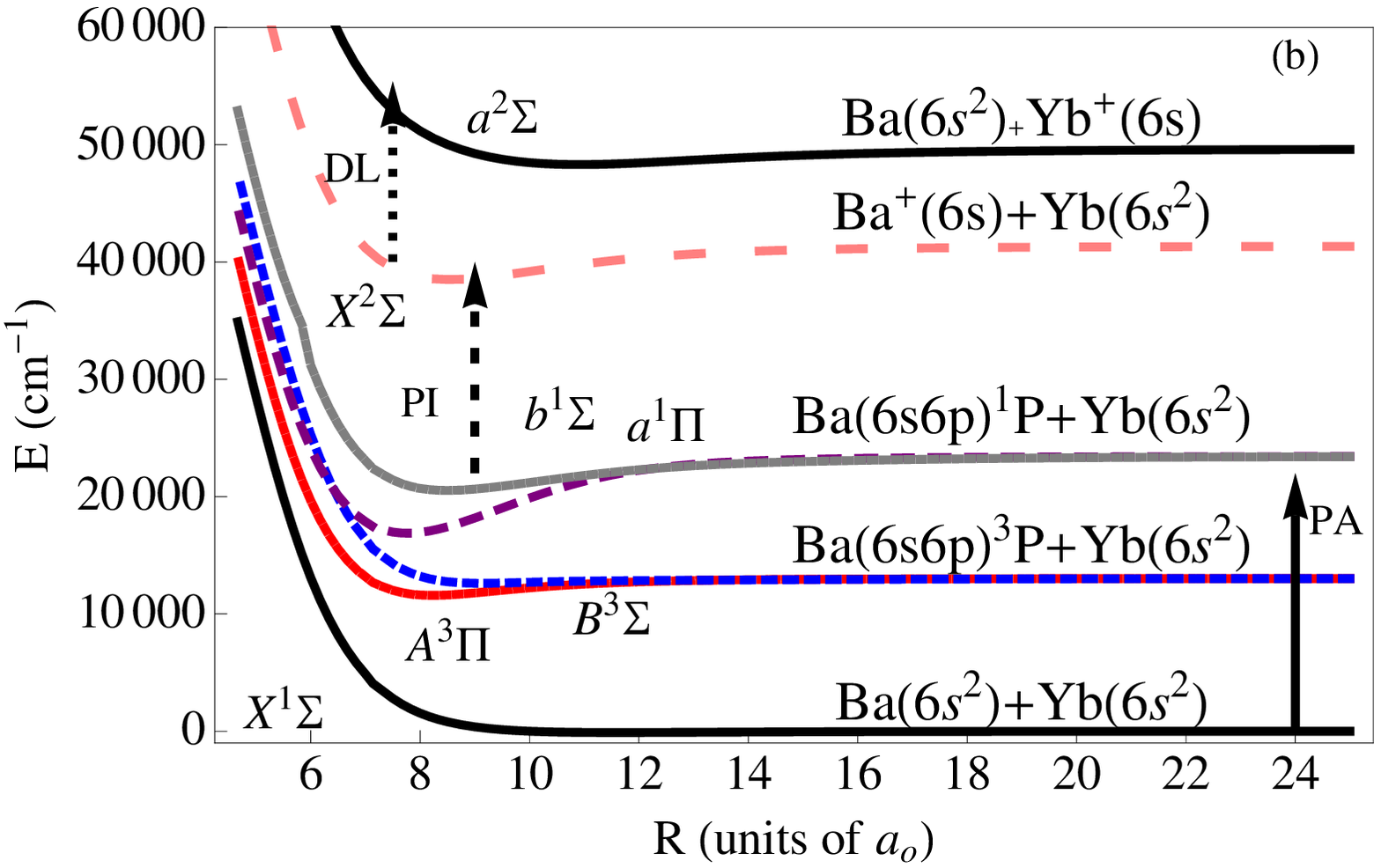}}
  \end{center}
  \caption{(Color online) Potential energy surfaces for the YbSr (a) and 
    YbBa (b) molecules in their neutral (lower set of curves) and ionic 
    (upper set of curves) states. $a_0$ is the Bohr length, $0.52918~\AA$. 
    The ground neutral state to the ground ionic state energy splitting in 
    the separated limit was optimized to produce the ionization levels known 
    experimentally. The arrows indicate the photo-association (solid arrow), 
    photo-ionization (medium dashed arrow) and proposed dissociation pulse 
    (short dashed arrow) in the electron EDM search.}
\end{figure}

We performed a SCF-HF followed by an MCSCF \cite{mcscf1,mcscf2} calculation 
on the neutral and ionic species of the alkaline earth-Yb separated atoms. 
Once again the Stuttgart basis set and ECPs ECP68MDF were used to describe 
the Yb \cite{stutYb} atom while the ECP36SDF and ECP54SDF\cite{stutRb2} 
basis set and ECP were used to describe the Sr and Ba atoms respectively. 
CPPs were included as well. In the neutral molecule we accounted for the 
X${}^1\Sigma$, A${}^3\Pi$, B${}^3\Sigma$, a${}^1\Pi$ and b${}^1\Sigma$ 
electronic states. In the ionic state we kept the three lowest ${}^2\Sigma$ 
states and the ${}^2\Pi$ state. The final step in the calculations involved 
performing an MRCI\cite{mrci1,mrci2} calculation on each spin multiplicity and 
group symmetry. The potential energy surfaces are presented in 
Figs.~\ref{fig3}~a and b. The upper (lower) set of curves in each panel are the 
ionic (neutral) molecules.

A possible route to make the molecular ions is presented in
Figs.~\ref{fig3} (a) and (b). By driving near the alkaline earth
${}^1S$~$\rightarrow$~${}^1P$ transition, one can photo-associate
molecules into a high-lying vibrational level of the ${}^1\Pi$ (or 
${}^1\Sigma$) state (solid arrow). From this state a photo-ionization pulse 
to the $X{}^2\Sigma$ state of the molecular ion should produce molecules in
the electronic state of interest for an electron EDM experiment (long dashed 
potential in the upper portion of each panel in Figs.~\ref{fig3}~(a) and (b)).

A typical Ramsey experiment would measure the energy difference
between two levels with differing angular momentum projections $M_s$
referred to a laboratory magnetic field.  The measurement consists
of the time-varying population of one of these states.  To measure
this, one could drive the molecule to the purely repulsive $a{}^2
\Sigma$ state (short dashed arrow), which dissociates into the Sr (Ba)
$5(6)\,s^2$ and Yb$^{\rm +}$ $6s$ atomic states. If the
dissociation laser were polarized to drive only one $M_s$ state,
then efficient detection of the Yb$^+$ ions serves as a high
signal-to-noise detection.

These potential energy surfaces of the alkaline earth-Yb ionic
molecules are markedly different from the neutral species alkali-Yb 
systems. First, the bond distance is reduced to around 8~a$_{\rm o}$ 
and the well depth increased by an order of magnitude. Physically, this 
is easy to understand. The Sr and Ba ions (in the separated limit) are 
very willing to accept an electron, hence they have a tendency to take an 
electron from the neutral Yb atom. Thus, charge is efficiently transferred 
from the Yb atom to the alkaline earth ion and an ionic-type bond is formed, 
as compared to the more shallow van der Waals-type  bond in YbRb and YbCs. 
This circumstance is also responsible for the larger body-fixed molecular 
electric dipole moment that in the neutrals.

The second surface in the ion is the $a{}^2\Sigma$ state, which is repulsive
above the equilibrium bond length of the $X{}^2\Sigma$. The excited state 
surface in YbBa$^{\rm +}$ is deeper than YbSr$^{\rm +}$, however the dissociation 
laser (DL) should still be able to drive the ground state to an unbound state 
of the system due to the rather large bond length of the $a{}^2\Sigma$ state 
in the molecular ions. The excited $a{}^2\Sigma$ state is more closely akin 
to the ungerade states of homonuclear molecules. In this instance, at the bond 
length of the $X{}^2\Sigma$ molecule, the configuration is a combination of two 
molecular orbitals, one of Ba(6s$^{\rm 1}$)-Yb(6s$^{\rm 2}$) and the other of 
Ba(6s$^{\rm 2}$)-Yb(6s$^{\rm 1}$) hybridized atomic configurations. The excited 
state is the orthogonal combination. Whereas the ground state has an electronic 
distribution favorable to a deep bond the excited state does not.

Table~\ref{tab4} gives the molecular parameters for a few states of
interest in the YbSr and YbBa systems. Once again, we fit the tabulated
points from the {\it ab initio} calculations to a Morse potential.
\begin{table}
  \caption{\label{tab4} Molecular parameters for YbSr and YbBa.
    Bond distance $r_e$, dissociation energy ($D$), vibration constants 
    ($\omega_e$ and $\omega_e\chi_e$), rotation constants ($B_e$ and $D_e$), 
    and vibration-rotation mixing constant $\alpha_e$ are all determined from 
    the Morse potential parameter. $r_e$ is in atomic units while all others are 
    in cm$^{\rm -1}$.}
  \begin{ruledtabular}
    \begin{tabular}{l|l|l|l|l|l|l|l}
      Molecule & $r_e$ & $D$ & $\omega_e$ & $\omega_e\chi_e$ & 
      $B_e$\,$10^{-2}$ & $D_e$\,$10^{-9}$ & $\alpha_e$\,$10^{-4}$ \\
      \hline
      YbSr $X{}^1\Sigma$ & 11.42 & 94 & 8.6 & 0.19 & 0.78 & 26 & 1.72\\
      YbSr $a{}^1\Pi$ & 7.30 & 8460 & 94.0 & 0.26 & 1.98 & 3.2 & 0.63\\
      YbSr$^{\rm +}$ $X{}^2\Sigma$ & 7.94 & 4230 & 61.5 & 0.22 & 1.62 & 4.5 
      & 0.70\\
      \hline
      YbBa $X{}^1\Sigma$ & 11.64 & 112 & 7.86 & 0.14 & 0.57 & 12 & 0.98\\
      YbBa $a{}^1\Pi$ & 7.67 & 6765 & 73.4 & 0.20 & 1.32 & 1.7 & 0.41\\
      YbBa$^{\rm +}$ $X{}^2\Sigma$ & 8.50 & 2810 & 42.8 & 0.16 & 1.08 & 2.7 
      & 0.47
    \end{tabular}
  \end{ruledtabular}
\end{table}
The neutral molecules contain very shallow wells. This is attributed to each 
atom influencing the other and causing a slight mixing of atomic $s$- and 
$p$-orbitals. This is a van der Waals interaction which creates the 
ground neutral molecular states. However, the molecular ion states are created 
by having an electron from Yb have significant amplitude at the site of 
the alkaline earth atom. There is expected to be significant mixing of 
atomic $s$- and $p$-orbitals on both atoms.

\section{Effective Electric Field Calculation}

The effective electric field is defined by the energy shift incurred by the 
EDM Hamiltonian 
$H_{\rm edm} = -{\vec d}_{\rm e}\cdot{\vec {\mathcal E}}_{\rm eff}$ 
acting on a putative electric dipole moment $d_{\rm e}$. We have previously 
detailed a perturbative method for estimating ${\mathcal E}_{\rm eff}$ to 
within $\sim 25\%$ based on our non-relativistic molecular 
orbitals\cite{meyer08,meyer06}, following the work in \cite{krippy}. 
The basic idea is that we can separate the calculation into an analytic 
and numeric piece via (in atomic units)
\begin{equation}
  {\mathcal E}_{\rm eff} = 
  \frac{\langle\Psi_{\rm mol}|H_{\rm edm}|\Psi_{\rm mol}
  \rangle}{-d_{\rm e}} = \frac{4\sigma}{\sqrt{3}}\epsilon_s\epsilon_p 
  \Gamma_{\rm rel} Z,
\end{equation} 
where $\Gamma_{\rm rel}$ is a relativistic factor that depends on the atomic 
properties and is given by an analytic expression. $\sigma=1/2$ is the 
electron projection on the molecular axis. $Z$ is the number of protons in the 
heavy atom. The values of $\epsilon_s$ and $\epsilon_p$ represent 
the amount of $s$- and $p$-atomic orbitals contained in the relevant molecular 
orbital, and are obtained from non-relativistic electronic structure 
calculations.

The estimate in the cases studied in \cite{meyer06,meyer08} was for systems 
with one heavy atom and one light atom. Therefore, the dominant contribution 
to ${\cal E}_{\rm eff}$ arises from the heavy atom. This is because the 
relativistic factor $\Gamma_{\rm rel}$ scales as $Z^2$ thus the effective 
electric field ${\mathcal E}_{\rm eff}$ scales as $Z^3$. In the examples we 
are considering here, both atoms contribute since they have similar $Z$. 
Therefore, it should be expected that the electron experiences an average 
effective electric field from both atoms. What remains to be determined 
is whether the atoms are polarized parallel or anti-parallel to each other; 
this will be addressed below.

Since $\Gamma_{\rm rel}$ depends on the atomic properties of the atom 
considered, we break the two contributions up by expanding our molecular 
wave function as follows:
\begin{eqnarray}\nonumber
  |\Psi_{\rm mol}\rangle &=& 
  \epsilon_{\rm s,A}|{\rm A}\;s\rangle + 
  \epsilon_{\rm p,A}|{\rm A}\;p\rangle + \\ && 
  \epsilon_{\rm s,Yb}|{\rm Yb}\;s\rangle + 
  \epsilon_{\rm p,Yb}|{\rm Yb}\;p\rangle,
\end{eqnarray}
and verified the condition $\langle\Psi_{\rm mol}|\Psi_{\rm mol}\rangle=1$ 
at all times in the calculation. The subscript A refers to either an alkali 
or alkaline earth atom. With this wave function we can calculate the quantity
${\mathcal E}_{\rm eff}$ via
\begin{eqnarray}\nonumber
  {\mathcal E}_{\rm eff} &=& 
  \frac{\langle\Psi_{\rm mol}|H_{\rm edm}|\Psi_{\rm mol}
  \rangle}{-d_{\rm e}}\\ &=& {\mathcal E}_{\rm eff,A} + 
       {\mathcal E}_{\rm eff,Yb}.
\end{eqnarray}
The only surviving terms are $\langle{\rm A,s}|H_{\rm edm}|{\rm A,p}\rangle$ 
and $\langle{\rm Yb,s}|H_{\rm edm}|{\rm Yb,p}\rangle$ and so the contributions 
from each atom simply add. The terms are individually calculated via the 
methods described in \cite{meyer08,meyer06}. Now, if the product 
$\epsilon_{s,{\rm A}}\epsilon_{p,{\rm A}}$ is of the same (opposite) sign as 
$\epsilon_{s,{\rm Yb}}\epsilon_{p,{\rm Yb}}$ then the individual effective 
electric fields ${\mathcal E}_{\rm eff,A}$ and ${\mathcal E}_{\rm eff,Yb}$ 
would add (subtract). Since each term has an uncertainty of $25\%$ associated 
with it, and because these errors are not completely random, we report that 
these current estimates are $50\%$ certain. 

One thing we discover in the {\it ab initio} calculations is that there are 
two large contributions to the overall molecular orbital, as discussed in 
Sec.~\ref{s:alkali}. In the alkalis, this mixing of molecular configurations 
is more pronounced than in the alkaline earths. These configurations are 
different zero angular momentum projections of atomic orbitals from each atom. 
We find that these two configurations are 
$|\Psi_1\rangle=|\sigma_2^1\sigma_3^0\rangle$ and
$|\Psi_2\rangle=|\sigma_2^0\sigma_3^1\rangle$ for two different molecular 
orbitals $\sigma_2$ and $\sigma_3$. 

In the case of YbRb and YbCs $\sigma_2$ is primarily a Rb (Cs) $5s$ ($6s$) 
atomic orbital with appreciable mixing of Yb $6s$ and $6p_z$ as well as Rb (Cs) 
$5p_z$ ($6p_z$) atomic orbitals. $\sigma_3$ has more of Rb (Cs) $5p_z$ 
($6p_z$) atomic character with larger admixtures of the other atomic orbitals. 
The YbSr$^+$ and YbBa$^+$ molecules also have similar discernment in 
the molecular orbital $\sigma_2$. The $\sigma_3$ molecular orbital is composed 
of the alkaline and Yb $p_z$ atomic orbitals with small admixtures 
of the respective $s$ atomic orbitals. 

If we make the approximation that these two configurations are all that 
matter in the calculation of ${\mathcal E}_{\rm eff}$, we can write the 
total molecular wave function as 
\begin{equation}
  |\Psi_{\rm tot}\rangle = c_1|\Psi_1\rangle + c_2|\Psi_2\rangle,
\end{equation}
where $c_1$ and $c_2$ are the MRCI coefficients. Ideally, $c_1^2 + c_2^2$ 
would be unity. However, since other configurations also contribute the sum 
$c_1^2 + c_2^2$ is approximately $0.95$ in all the cases considered.

The total $s$ and $p$ contribution from one of the atoms is given by the 
weighted sum (weighted by the MRCI coefficients) of the individual $s$- 
and $p$-atomic orbitals as follows:
\begin{equation}
  |\psi_{\rm A,s}\rangle = 
  c_1\sum_{j=1}^k\alpha_j g_j + c_2\sum_{j=1}^k\beta_j g_j.
\end{equation}
In the above $g_j$ is a Gaussian centered on atom A having $s$-wave 
characteristics. The coefficients $\alpha_j$ and $\beta_j$ describe the 
relative contributions of each Gaussian $g_j$ to the molecular orbital. 
$c_1$ and $c_2$ are the same MRCI coefficients from the molecular orbital 
calculation. We can now define the $\epsilon_{s,{\rm A}}$ from atom A 
as\cite{meyer06,meyer08}
\begin{equation}
  \epsilon_{s,{\rm A}} = \frac{\langle \psi_{\rm A,s}|\Psi_{\rm tot}\rangle}
          {\langle \psi_{\rm A,s}|\psi_{\rm A,s}\rangle}.
\end{equation}
Similar definitions describe $\epsilon_{p,{\rm A}}$, $\epsilon_{s,{\rm Yb}}$, and
$\epsilon_{p,{\rm Yb}}$. The results of the calculation are presented in 
Table~\ref{tab5}.

\begin{table}
  \caption{\label{tab5} Effective electric field estimates for YbRb, YbCs,
  YbSr$^{\rm +}$, and YbBa$^{\rm +}$ in GV/cm. Also presented are
  the molecular electric dipole moments (Debye) and the critical field for
  polarizing the molecules (kV/cm).}
  \begin{ruledtabular}
    \begin{tabular}{l|l|l|l|l|l}
      Molecule & ${\mathcal E}_{\rm eff}$ & ${\mathcal E}_{\rm eff}$(A) & 
      ${\mathcal E}_{\rm eff}$(Yb) & $d_{\rm m}$ & ${\mathcal E}_{\rm pol}$\\
      \hline
      YbRb $X{}^2\Sigma$ & -0.70 & 0.45 & -1.15 & 0.21 & 5.5\\
      YbCs $X{}^2\Sigma$ & 0.54 & 1.42 & -0.88 & 0.24 & 3.5\\
      YbSr$^{\rm +}$ $X{}^2\Sigma$ & -11.3 & 10.6 & -21.9 & 5.1 & 0.38\\
      YbBa$^{\rm +}$ $X{}^2\Sigma$ & 1.2 & 12.6 & -11.4 & 5.1 & 0.25
    \end{tabular}
  \end{ruledtabular}
\end{table}

In Table~\ref{tab5} we have separated out the contributions from both atoms 
individually. It is apparent that both atoms contribute to the size 
of the effective electric field, but with the opposite sign. We interpret 
this as both atoms polarizing in different directions with respect to the 
molecular axis. Indeed, this is what is observed when one looks at the 
electronic distribution of the unpaired electron. We also note that 
${\mathcal E}_{\rm eff}$ of the YbRb and YbCs systems makes them an attractive 
alternative to experiments on the solitary alkali atoms. Even though 
${\mathcal E}_{\rm eff}$ is small compared to other $X{}^2\Sigma$ molecules, 
the expected long coherence times in an ultracold sample still makes them 
appealing. For the lone electron in Rb (Cs) atomic electron 
EDM experiments one would have to apply an electric field ${\mathcal E}_{\rm app}$ 
of 27~MV/cm (4.5 MV/cm) --- using the enhancement factors in \cite{rbedm} --- 
to achieve the same ${\mathcal E}_{\rm eff}$. This is to be contrasted 
with the proposed alkali-Yb molecular electron EDM experiment where one needs 
${\mathcal E}_{\rm app}\approx$10--25~kV/cm. The values of 
${\mathcal E}_{\rm app}$ in the molecular cases are based on multiplying 
${\mathcal E}_{\rm pol}$ by 3--5 in order to be in the linear Stark regime. 
${\mathcal E}_{\rm pol}$ is defined as the electric field required to equalize 
the Stark energy and the energy splitting between rotational levels in the 
molecule. ${\mathcal E}_{\rm pol}$ is tabulated in Table~\ref{tab5}.

We have used the following convention for the overall sign of 
${\mathcal E}_{\rm eff}$: we define the molecular axis ${\hat n}$ as pointing 
from Yb to A, where A is either an alkali or alkaline earth atom. This is 
because we choose the axis to point from more negative charge to more positive. 
In the alkali-Yb molecules, Yb has negative charge while in the alkaline 
earth-Yb molecules it has less positive charge. A positive value of 
${\mathcal E}_{\rm eff}$ means that it lies against the direction of ${\hat n}$, 
i.e. points from positive charge toward negative charge. Positive 
${\mathcal E}_{\rm eff}$, in turn, means that the atomic electron 
is polarized so that it is displaced in the direction $-{\hat n}$. 

These molecules offer alternatives to performing the experiments on Rb and Cs
or Sr$^{\rm +}$ and Ba$^{\rm +}$ alone. In the atoms, the size of 
${\mathcal E}_{\rm eff}$ is proportional to the applied field 
${\mathcal E}_{\rm app}$ by an enhancement factor\cite{krippy}. In molecules, 
this is not the case provided they are fully polarized. However, polarizing
these molecules is not as easy to do as in the case of ${}^3\Delta_1$ molecules
discussed in \cite{meyer08,meyer06}. Yet, the ${\mathcal E}_{\rm eff}$
that can be attained is larger than the ${\mathcal E}_{\rm eff}$ that can 
be attained in the atomic systems for a similar ${\mathcal E}_{\rm app}$
\cite{krippy}. Therefore, with the recent advances in producing cold
molecules, these systems offer very intriguing and viable alternative routes 
to probe for the electron EDM.

The idea of creating ultracold molecules from ultracold atoms, along with 
the anticipated benefit of molecules to electron EDM searches, to explore 
a variety of laser cooled atoms paired with Yb for an electron EDM experiment 
is the focus of this paper. We chose heavier pairing partners in the hope of 
reducing the rotational constant of the molecules, thereby making them easier 
to polarize in modest applied electric fields. However, as one increases the 
mass of the pairing partner, one also increases the contribution it gives to 
${\mathcal E}_{\rm eff}$ and this contribution is always opposite to Yb's 
contribution, thereby reducing the effect. Therefore, there should be some 
trade off between reduced rotational constant and overall 
${\mathcal E}_{\rm eff}$. Future work would include exploring systems such as 
YbK, YbNa, YbCa$^+$, and YbMg$^+$. In addition, the same ideas would work for 
Hg-paired partners. Work on HgH\cite{HgH} has already been performed in regards 
to an electron EDM experiment. Systems such as HgLi, HgNa, etc. should also 
give similar results as the Yb-paired systems with the added benefit that Hg 
is heavier than Yb, thereby hopefully producing a larger ${\mathcal E}_{\rm eff}$.

\section{Conclusions}

We have considered a particular class of ${}^2\Sigma$ molecules that offer some 
advantages and drawbacks to current polar molecule electron EDM searches. 
This type of molecule is formed from ultracold atoms so that 
the ground state molecules are also ultracold, leading to the 
possibility of longer coherence times. We have considered YbRb (-0.7~GV/cm), 
YbCs (+0.54~GV/cm), YbSr$^+$ (-11.3~GV/cm) and YbBa$^+$ (+1.2~GV/cm). 

YbRb and YbCs can be polarized in fairly modest fields and thereby are 
intriguing inasmuch as they offer some incentive over searching for the 
electron EDM in the alkali atom alone. The molecular ions YbSr$^+$ and 
YbBa$^+$ have many useful attributes, However, the field required to polarize 
these molecules is rather large for an ion trap. Thus, the use of an optical 
lattice may be more practical for holding these ions for a long period of 
time without having to rotate an ${\mathcal E}_{\rm app}$ on the order of 
1~kV/cm.

\begin{acknowledgments}
We wish to acknowledge D. DeMille for bringing Yb-containing molecules to our 
attention, an anonymous referee for a crucial observation, and the NSF for 
financial support.
\end{acknowledgments}

\end{document}